# Closing the Prior-Posterior Loop: Self-Reflective Molecular Design with Analysis-Driven LLM Iteration

Junyi GONG[a], Zijie QIU[b], Ben Zhong TANG*[b, c]

a. Faculty of Chemistry, Shenzhen MSU-BIT University

b. School of Science and Engineering, Chinese University of Hong Kong (Shenzhen)

c. Department of Chemistry, Hong Kong University of Science and Technology

## Abstract

Can a general-purpose large language model design molecules with the precision of a seasoned chemist? Current LLM-based frameworks answer this question with scalar feedback loops—generate, score, reject—that amount to informed trial-and-error. Here we show that replacing a single number with the full physicochemical rationale from first-principles calculations transforms the LLM from a stochastic sampler into a causal reasoner. Our system couples retrieval-augmented generation with a self-reflection module that feeds orbital energies, atomic charges, and electron densities—rather than compressed scores—back into the design loop. On HOMO-LUMO gap targets from 2.0 to 5.0 eV, this structure-property-relationship (SPR) reflection achieves a deviation as low as 0.0014 eV with a 100% success rate under the SPR+RAG configuration, consistently outperforming scalar-feedback and non-reflective baselines in median and mean deviation. The framework generalizes seamlessly to dipole-moment design, synthetic accessibility optimization, and molecular docking, and proves robust across 7 distinct LLM backbones. These results establish a new paradigm: when the model understands not only that a molecule fails, but why, iterative molecular design becomes genuinely mechanistic.

## Introduction

The design of novel molecules endowed with specific properties stands as a paramount priority in the field of chemical science. Yet, for centuries, this endeavor more closely resembled a craft—reliant upon intuition, trial, and accumulated experience—rather than a rigorous, systematic, and predictive scientific technique. Especially within the domain of organic systems, despite the availability of well-established theoretical frameworks and quantum mechanical methods, the inherently emergent phenomena give rise to design principles that remain fundamentally unpredictable and deeply reliant on empirical heuristics.[1]

Despite the availability of a priori knowledge from scientific literature and datasets, molecular design remains fundamentally constrained by the gap between theoretical understanding and practical realization. Conventional machine learning systems can generate molecules conditioned on prior knowledge, yet they lack the capacity to extract causal insights from their own generative outputs—post hoc physical-chemical rationales that are essential for mechanistic refinement. This imposes a fundamental limitation: such systems are capable of correlational learning but not of causal learning from generation.[2–12]

In recent years, large language models (LLMs) have demonstrated remarkable cross-disciplinary capabilities, yielding promising results in the formal sciences—including mathematics, logic, and computer science.[13–15] Within the field of chemistry, LLMs have likewise been introduced to facilitate the design of molecules and materials endowed with specific properties. While recent LLM-based frameworks, such as ChatDrug and TaLiRAGen, have commendably introduced iterative paradigms with domain feedback, the nature of this feedback often limits the depth of iteration.[16–21] In these systems, the evaluator typically provides scalar feedback—such as numerical scores, binary accept/reject signals, or pre-digested summaries. Although such outcome-driven feedback effectively filters candidates, it inadvertently compresses the rich physicochemical landscape into a single metric. Consequently, the LLM is informed whether a molecule fails, but not why it fails, rendering the iterative process more akin to a heuristic search than a mechanism-driven refinement.

What remains critically absent is computation-grounded reflection—a process that upgrades iterative feedback from the outcome level to the mechanism level. This demands that the LLM interpret raw computational outputs—such as orbital energies, electron density distributions, and bond parameters—derived from first-principles calculations, thereby understanding why a molecule fails to satisfy target properties and how it can be systematically improved, rather than merely receiving numerical feedback. Such reflection transforms iterative design from heuristic exploration into causal reasoning. Here, 'causal' refers to the LLM's capacity to induce structure-property relationships from paired molecular structures and their computed descriptors—an operationally mechanistic inference that links specific substructural motifs to changes in electronic properties. This stands in contrast to the opaque correlation learned by conventional predictive models.

In this work, we introduce the Framework for Organic-molecule Generation via Reflection-Guided Evaluation (FORGE, Table 1), a molecular design system that bridges the a priori-scientific literature; a multi-stage evaluation protocol—combining semi-empirical prescreening with first-principles computation—that grants direct access to raw computational outputs beyond superficial scores; and a self-reflection module that performs causal analysis of computational results to guide iterative improvement. Unlike existing systems that treat computation as a black-box scoring function, our approach views computation as a rich repository of physicochemical information—one that enables genuine learning from generation.

| Work | Type | Target | Reported Metrics | Precision / Accuracy | Comparison with FORGE | Ref |
|---|---|---|---|---|---|---|
| ChatDrug | LLM molecule generator | Tanimoto similarity | Success rate across 39 tasks | No deviation metrics | Different application domain | [16] |
| Bhattacharya | LLM molecule generator | Electronic structure adjustment | Success rate | No deviation metrics | No reflection mechanism | [22] |
| Vogiatzis | LLM molecule generator | $CO^2$-philicity | DFT-based evaluation | No target deviations | Different application domain | [23] |
| Ito et al. | LLM molecule generator | Organic structure-directing agents | Iterative refinement | No deviation metrics | Different application domain | [18] |
| TaLiRAGen | LLM molecule generator | Docking score | Docking performance | No deviation metrics | Different application domain | [17] |
| **FORGE (This work)** | **LLM molecule generator** | **QM properties** | **Median dev. down to 0.0014 eV** | **Continuous deviation metrics** | - | - |
| GNN predictors (e.g. SchNet, DimeNet) | Property predictor | HOMO-LUMO gap | Prediction MAE: 0.06-0.19 eV | - | Predictor, not generator | [24,25] |
| Conditional generators (e.g. cG-SchNet) | Deep learning molecule generator | - | Validity, novelty, diversity | - | Not focused on target accuracy | [26] |

*Table 1: Comparison of the FORGE paradigm with reported similar work based on LLM or neural network.*

# Methodology

## Overall framework: Analysis-driven iteration

A mature researcher working in a specialized domain of molecular design—that is, engineering molecules with targeted properties—typically follows a traditional paradigm: learn, realize, design, evaluate, and reflect. In the learn phase, the researcher first absorbs design principles from existing literature, theories, and datasets, then abstracts and distills common features and underlying rules. Once these patterns are realized, the researcher proceeds to design candidate molecules that embody them. The evaluate step may rely on computational chemistry or wet experiments to determine whether the molecule meets the specified requirements. If it does, the molecule can be deployed; if not, the researcher must diagnose the shortcomings, reflect on the feedback, and incorporate those insights into a new design cycle. Our objective is to employ an LLM as a surrogate for the human researcher, thereby automating and accelerating this iterative process.

Our system operates on a novel iteration paradigm: Generate → Analyze → Reflect → Refine. Unlike traditional “Generate → Score → Regenerate” loops, which rely on scalar outcomes, our loop incorporates mechanism-level analysis of computation results to inform reflection.

## Architecture

The system comprises three main components: a retrieval-augmented generation (RAG) module, an LLM core, and a reflection module. These components operate in concert with the core LLM to form a complete molecular design framework.

The general procedure of the system is as follows (Figure 1):

1. The target properties of the molecule—such as HOMO-LUMO gap, dipole moment, or other in silico computable descriptors—are specified as input.
2. Relevant a priori knowledge (e.g., scientific papers, datasets) is compiled and organized into a RAG database.
3. The system retrieves pertinent a priori knowledge from the database on the basis of the target properties.
4. The LLM generates an initial candidate molecule informed by the retrieved knowledge.
5. The caculation system in the reflection module calculates the raw computational outputs of the candidate molecule.
6. The evaluator will assess the candidate molecule against the target properties.
7. If a generated molecule does not meet the required criteria, the system analyzes the outputs using a self-reflection mechanism to extract key physicochemical parameters. The retrieval-augmented generation (RAG) module is then updated with this reflection information, which the LLM incorporates into the next design iteration. This iterative process continues until the candidate molecule satisfies all specified target properties.

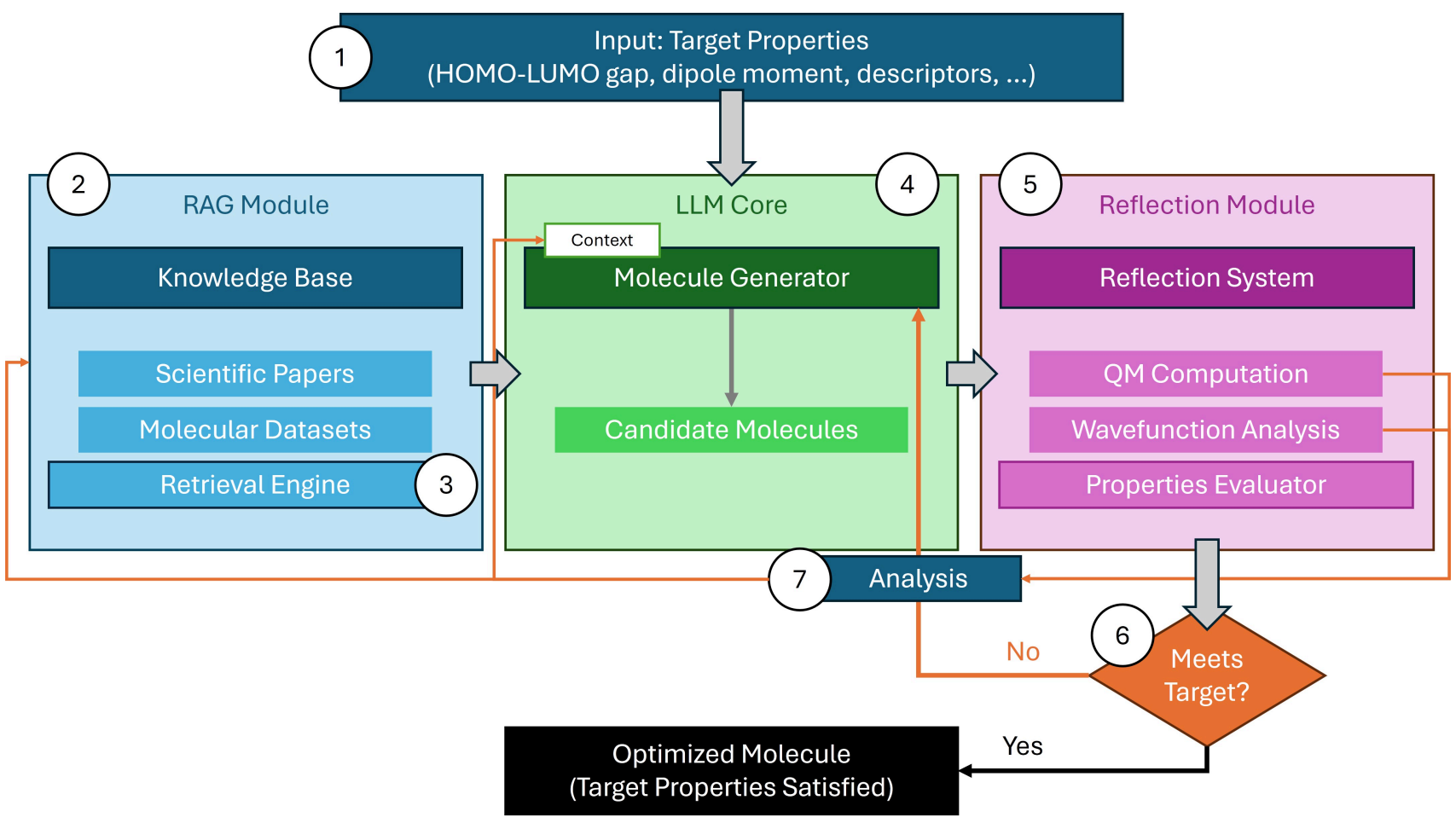

*Figure 1: The architecture of the autonomous molecular design system.*

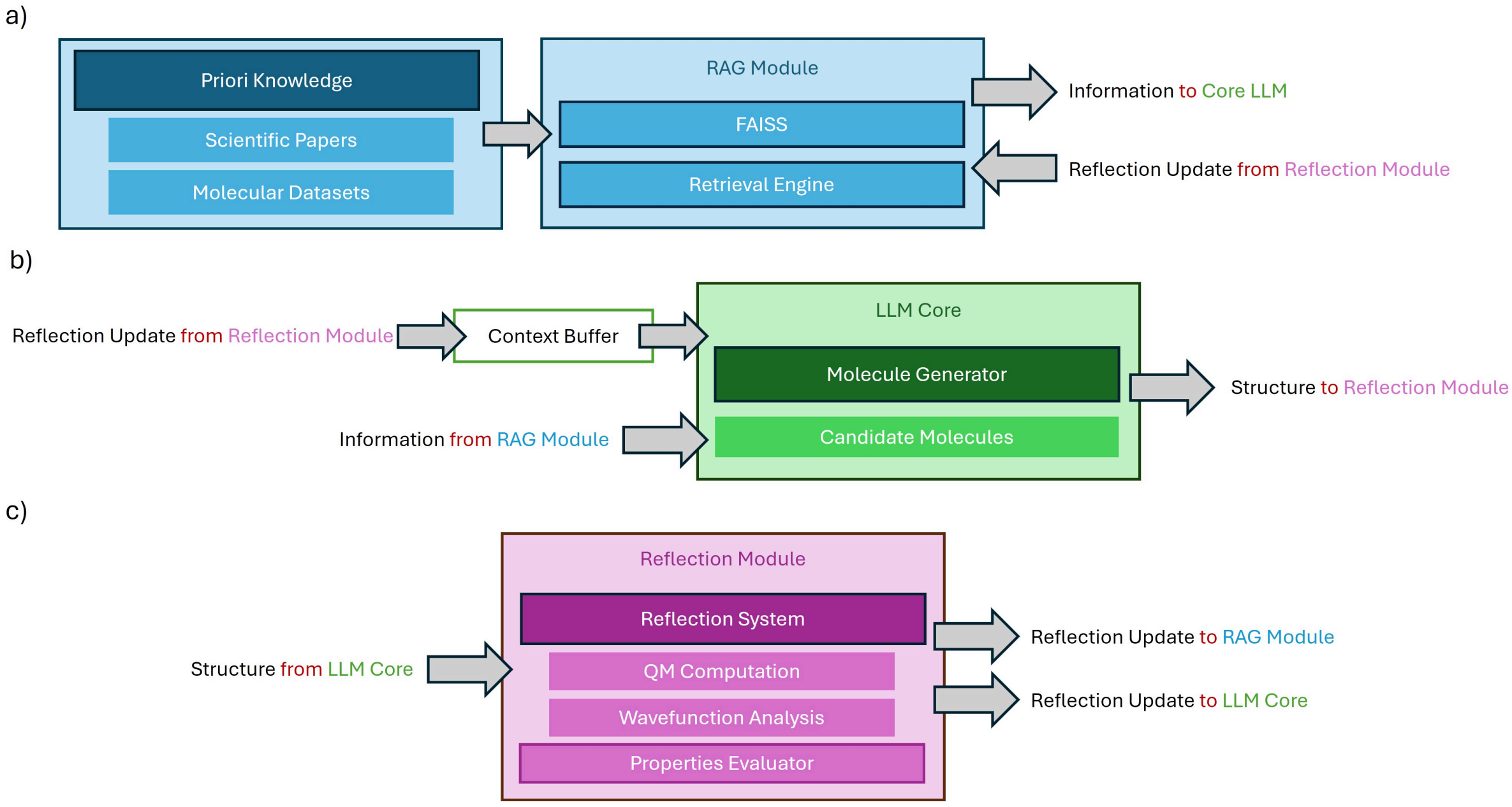

*Figure 2: The information flow of the three components: RAG (a), LLM core (b), and reflection module (c).*

The Figure 2 figure illustrates the information flow between the three components of the system.

### The RAG module

The RAG module is responsible for retrieving pertinent prior knowledge from the database on the basis of the target properties, and for pushing relevant information to the LLM core for reference. The RAG can be updated with feedback from the reflection module, rendering the overall system more dynamic and adaptive.

Ideally, it could be formed by priori knowledge of the target properties like literature, datasets, or theories. In our implementation, the knowledge was built on QM9 dataset with FAISS vector database, which contains 130,000 organic molecules which contains fewer than 9 heavy atoms, with their computed properties.

### The LLM core

The LLM core is responsible for generating candidate molecules informed by the retrieved knowledge. It also receives reflection information through a context buffer, thereby implementing a self-reflection mechanism.

### The reflection module

The reflection module is designed not merely to invoke evaluations on structures from the LLM core, but to provide more information to the LLM and to update the RAG module. In a task such as generating molecules with a specific HOMO-LUMO gap or dipole moment, these outputs may include:

- Orbital energy levels and spatial distributions
- Population analysis
- Wavefunction information

Unlike conventional protocols that return only the target properties as scalar values or scores, our reflection module preserves the rich physicochemical information contained in DFT outputs. Furthermore, the evaluation process can determine the speed of the system; thus, performance optimization is required.

The self-reflection mechanism transforms raw DFT outputs into actionable insights through a three-step process:

- Step 1: Information Extraction — Parse DFT outputs to extract key physicochemical parameters.
- Step 2: Causal Reasoning — Identify relationships between molecular structure and target properties.
- Step 3: Strategic Planning — Propose specific molecular modifications based on causal analysis.

This structure-property relationship (SPR) reflection mechanism not only extracts the final target properties but also captures physicochemical auxiliary information that explains why a given structure yields those properties. This enables the LLM to construct a causal structure-property relationship, thereby serving as the most critical component of the system.

### The evaluation process

In this work, to improve efficiency, the quantum-mechanical (QM) calculations within the reflection module are partitioned into multiple stages. First, the evaluator receives a series of SMILES strings from the LLM core and generates their initial 3D structures using the MMFF method implemented in RDKit. These structures are then optimized with GFN2-xTB, which also computes their semi-empirical properties. Next, the optimized chemical structures and their xTB-derived properties are fed into a multilayer perceptron regressor (MLPR) that has been pretrained to approximate DFT-level properties from xTB data, with chemical structures encoded as MACCS keys. The MLPR maps the xTB properties to calibrated DFT-equivalent values. Molecules are then ranked according to the deviation between these calibrated properties and the target, and the top candidates are selected for further DFT calculations using PySCF. Together, these steps constitute a multi-stage evaluation pipeline (Figure 3) that balances computational efficiency with predictive accuracy.

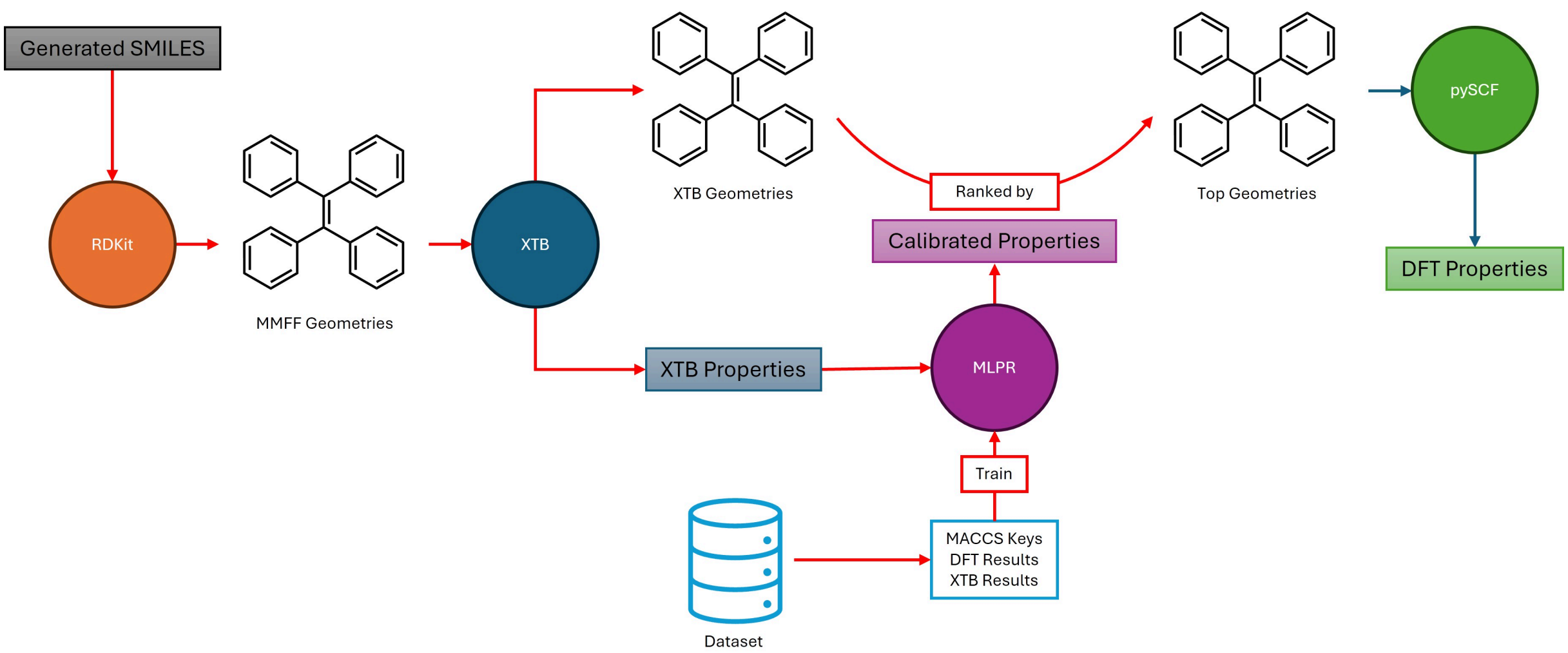

*Figure 3: The architecture of the multi-step evaluation process.*

The multi-step evaluation process can improve the overall performance of the system. In this batch process, the system receives x candidates from the LLM core, performs a fast evaluation, and then selects the top y candidates for accurate evaluation. For example, when the target property is the HOMO-LUMO gap or the dipole moment, we employ GFN2-xTB to optimize all geometries and screen for the desired properties, then select a subset of molecules for high-precision DFT calculations. The resulting outputs are subsequently evaluated by the evaluator and analyzed through the self-reflection mechanism. The LLM incorporates these insights into the generation of the next batch of candidate molecules. This iterative process continues until the candidate molecules meet the specified target properties. Unless otherwise specified, all experiments were conducted using a batch-reflection approach, in which the LLM first generated 20 candidate molecules for pre-screening; the top 5 candidates were then subjected to an accurate evaluation ($x = 20$, $y = 5$). Alternatively, a per-molecule reflection strategy is also possible. This is equivalent to batch reflection with $x = 1$ and $y = 1$, focusing on a single candidate molecule for evaluation and reflection.

Theoretically, batch reflection and per-molecule reflection represent two viable strategies, neither of which is categorically superior. The choice between them is determined by the specific task and the availability of computational and LLM resources. Per-molecule reflection entails a higher density of LLM reasoning, whereas batch reflection intrinsically combines scalar-reflection and SPR reflection, thereby aggregating a greater volume of reflective information across candidates. We further discuss the trade-offs between the two approaches in the Results section.

# Results

## The priori-posteriori loop

The central premise of our implementation is to combine a priori and a posteriori knowledge in the generation of molecules. In recent years, LLM-based molecular generators have been developed primarily from the perspective of a priori knowledge, typically drawn from existing datasets or the scientific literature. However, these systems employ the evaluator merely as a provider of outcome-level scores rather than as a source of mechanistic reasoning, and therefore rarely evolve on the basis of causal a posteriori insights derived from their own generative outputs.

Our system addresses this limitation by integrating a RAG module, an LLM core, and a reflection module. The RAG module retrieves pertinent a priori knowledge from the database on the basis of the target properties. The reflection module encapsulates the evaluation criteria (e.g., DFT calculation software), and the self-reflection mechanism allows the LLM to reflect by analyzing its own generative outputs, find why the generated molecule does not meet the requirements, and adjust its behavior accordingly.

### Iterations and convergence

Before conducting the experiments, it was necessary to determine the number of iterations required to achieve the desired performance. To this end, we first designed an experiment consisting of three parallel runs, each initiated with distinct random seeds and comprising ten iterations of tasks sharing identical target properties. The performance of the system was monitored throughout. We observed that system performance improved when the number of iterations exceeded three.

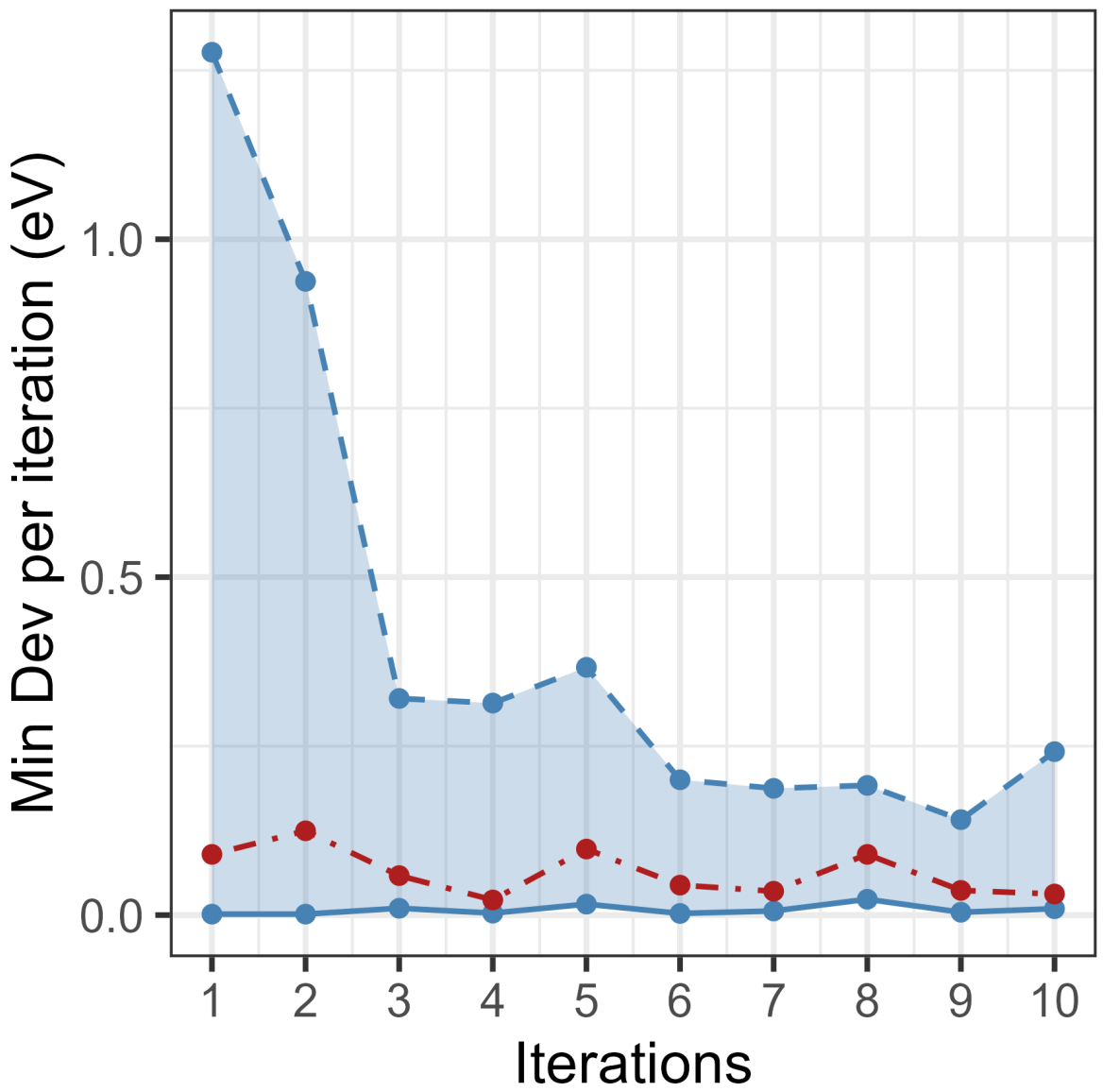

*Figure 4: Minimum deviation range per iteration as a function of iteration number in accumulated experimental data for generating molecules with target HOMO-LUMO gaps of 3.0, 4.0, and 5.0 eV.*

The results are presented in Figure 4. They reveal that the convergence behaviour of the SPR model is not strictly monotonic. In each iteration, the extreme performance (minimum, blue solid line) does not necessarily improve and may, in some cases, even oscillate. This may reflect an inherent bound on the reasoning capability of the specific large language model. The typical performance (median, red dashed line) tends to oscillate upon encountering a local minimum, possibly as a consequence of overthinking. The worst performance (blue dashed line) exhibited a generally convergent behaviour.

Therefore, a prudent monitoring strategy involves conducting several (e.g., 3-5) independent runs with a moderate number of iterations (e.g., 5-10), recording all data, and subsequently selecting the best outcome after all experiments are completed.

**Ablation study with different difficulties**

To evaluate the capability of the system, we conducted a series of experiments on designing molecules with a specific HOMO-LUMO gap. Although this is a fundamental property, designing a molecule with a precisely targeted HOMO-LUMO gap remains a challenging task for a human researcher. A human researcher may excel at designing a system that is more red-shifted or blue-shifted, but not at achieving an exact HOMO-LUMO gap. Nevertheless, precise control over the HOMO-LUMO gap is of considerable significance in materials discovery. Moreover, the HOMO-LUMO gap is one of the basic properties in the QM9 dataset, which serves as the foundation for our RAG implementation in this work. In these experiments, the reflection step ultimately relies on the pySCF package to perform DFT calculations. To emphasize the importance of reflection, we also implemented two distinct reflection variants: scalar-reflection, in which the HOMO-LUMO gap is the sole piece of information fed back to the LLM; and SPR reflection, which additionally provides the HOMO and LUMO energies, Mulliken atomic charges, total electronic energies, and dipole moments from the DFT outputs.

All experiments were performed using DeepSeek-V4Pro unless otherwise noted, as this model strikes an optimal balance between reliability, and performance.

We evaluated five target HOMO-LUMO gaps: 5.0, 4.0, 3.0, 2.0, and 1.0 eV. From a chemical perspective, designing a molecule with a precise HOMO-LUMO gap of 1.0 eV is the most challenging task because it is the smallest gap. The 5.0 eV and 4.0 eV targets are comparatively easier, as they fall within the most typical energy-gap range in the QM9 dataset, which serves as the foundation for our RAG implementation in this work. The 3.0 eV and 2.0 eV gaps (the typical range of organic fluorescence agents) present moderate difficulties, requiring greater design effort and computational reasoning.

The evaluation of LLM performance follows specific principles. Owing to the inherent stochasticity of the reasoning process, an ablation study—performing independent runs with different seeds and then comparing the statistical outcomes—is essential to ensure the reliability and validity of the results. The first metric is the success rate, defined as the proportion of runs that yield meaningful molecules; this metric captures the stability of the system. Performance can be measured by the minimum, median, and mean deviations, which represent the best-case, typical, and average performance, respectively. For optoelectronic applications, the highest deviation could be tolerated is assumed to be 0.10 eV. Consequently, if a run yields a deviation exceeding this threshold, its significance remains marginal, even if it is counted as a successful generation.

| Mode | Target (eV) | Success Rate | Min Dev (eV) | Median Dev (eV) | Mean±SE (eV) | LLM Time (s) | Tokens |
|---|---|---|---|---|---|---|---|
| SPR+RAG | 5.0 | 3/3 | 0.0094 | **0.0203** | **0.0174±0.0040** | 706 | 71,158 |
| | 4.0 | 3/3 | **0.0014** | **0.0014** | **0.0023±0.0009** | 635 | 67,462 |
| | 3.0 | 3/3 | 0.0054 | **0.0061** | **0.0068±0.0021** | 789 | 76,941 |
| | 2.0 | 3/3 | 0.0129 | **0.0134** | **0.0178±0.0047** | 1168 | 92,189 |
| Scalar+RAG | 5.0 | 3/3 | 0.0203 | 0.0224 | 0.0217±0.0007 | 986 | 76,437 |
| | 4.0 | 3/3 | 0.0024 | 0.0044 | 0.0054±0.0021 | 882 | 71,617 |
| | 3.0 | 3/3 | **0.0032** | 0.0108 | 0.0117±0.0052 | 825 | 78,557 |
| | 2.0 | 3/3 | **0.0023** | 0.0632 | 0.0562±0.0293 | 1129 | 96,408 |
| RAG-only | 5.0 | 3/3 | 0.0160 | 0.0224 | 0.0204±0.0020 | 935 | 63,208 |
| | 4.0 | 3/3 | 0.0014 | 0.0101 | 0.0076±0.0031 | 756 | 57,998 |
| | 3.0 | 3/3 | 0.0858 | 0.1256 | 0.1123±0.0133 | 1216 | 77,553 |
| | 2.0 | 3/3 | 0.1775 | 0.2734 | 0.2609±0.0450 | 1084 | 80,330 |
| Baseline | 5.0 | 3/3 | **0.0078** | 0.0330 | 0.0371±0.0182 | **303** | **14,435** |
| | 4.0 | 3/3 | 0.0351 | 0.0828 | 0.0718±0.0189 | **297** | **17,736** |
| | 3.0 | 3/3 | 0.0034 | 0.0065 | 0.0303±0.0254 | **387** | **26,980** |
| | 2.0 | 3/3 | 0.1547 | 0.1549 | 0.1548±0.0001 | **1086** | **69,071** |

*Table 2: Ablation study on DFT-based HOMO-LUMO gap design: deviations across 4 target values (5.0, 4.0, 3.0, 2.0 eV) and four configurations.*

| Mode | Target (eV) | Success Rate | Min Dev (eV) | Median Dev (eV) | Mean±SE (eV) | LLM Time (s) | Tokens |
|---|---|---|---|---|---|---|---|
| SPR+RAG | 5.0 | 3/3 | 0.0017 | 0.0017 | 0.0019±0.0002 | 561 | 41,791 |
| | 4.0 | 3/3 | 0.0013 | 0.0027 | 0.0051±0.0031 | 515 | 41,148 |
| | 3.0 | 3/3 | 0.0012 | 0.0063 | 0.0067±0.0033 | 534 | 44,081 |
| | 2.0 | 3/3 | 0.0002 | 0.0003 | 0.0004±0.0002 | 605 | 47,645 |

*Table 3: Ablation study on GFN2-XTB-based HOMO-LUMO gap design: deviations across 4 target values (5.0, 4.0, 3.0, 2.0 eV) in the SPR+RAG approach.*

Deviations can serve markedly different roles across research fields. In quantum chemistry calculations, standard DFT methods typically exhibit deviations of 0.2-0.5 eV when benchmarked against high-precision post-HF methods. In optoelectronic systems, a deviation of 0.1 eV is commonly regarded as the threshold for practical applicability. For instance, in solar cell applications, a deviation of 0.1 eV or less is generally considered acceptable for commercialization. By contrast, in fields such as catalyst design, a deviation of 0.05 eV or less may be required to achieve the desired catalytic activity.

The ablation study results (shown in Table 2) demonstrate the high potential of the FORGE paradigm (SPR+RAG). In routine tasks, such as predicting the HOMO-LUMO gap in the range of 3.0 eV to 5.0 eV, it typically exhibits a promising performance, with errors still below 0.01 eV. For more challenging tasks, such as the HOMO-LUMO gap of 2.0 eV, the typical error remains within 0.1 eV. The reference target is computed at the DFT level (B3LYP/6-31G(2d,p)), which is identical to that used in the QM9 dataset. This raises an important question: Does FORGE genuinely reason through reflection, or does it merely perform interpolation or extrapolation?

To answer this question, we performed an additional reference experiment in which the DFT evaluation component was replaced by pure GFN2-xTB properties, so that the LLM reasons directly on xTB outputs without the MLPR calibration step. As shown in Table 3, the SPR+RAG approach exhibits only subtle deviations even in more challenging tasks, confirming that the reflection and reasoning processes are indeed effective—the LLM is genuinely reasoning rather than merely retrieving or interpolating.

Notably, for the difficult 2.0 eV target, the pure xTB evaluator yields a median deviation of only 0.0002 eV, in stark contrast to the 0.0129 eV observed under the DFT-based pipeline. However, this result should not be interpreted as evidence that xTB provides superior accuracy. Semi-empirical xTB methods are inherently less accurate than DFT for narrow-gap systems, which typically involve extended $\pi$-conjugation or strong intramolecular charge-transfer characteristics. The low deviation under xTB reflects the internal consistency of the evaluator rather than its absolute fidelity. In the DFT-based pipeline, the MLPR calibrator—trained to map xTB properties to DFT-equivalent values within the QM9 domain—struggles to generalize to molecules with more than nine heavy atoms. This out-of-distribution (OOD) limitation introduces calibration noise that degrades the feedback provided to the LLM, ultimately inflating the observed deviation.

This comparison illuminates a fundamental trade-off in multi-stage evaluation pipelines: the xTB evaluator offers a self-consistent and computationally inexpensive feedback signal, but at the cost of absolute accuracy; the DFT-based pipeline aims for higher fidelity, yet its reliance on an ML surrogate for pre-screening introduces OOD vulnerability that can under-mine the final result. Critically, these findings underscore that the design of the evaluator is a decisive factor in the overall system performance. An ideal evaluator must balance computational efficiency with reliable generalization to the chemical space explored during generation. Our results further reveal a subtle yet crucial insight for closed-loop optimization: the internal self-consistency of the evaluator can outweigh its absolute accuracy in guiding iterative refinement. A method like GFN2-xTB, despite systematic deviations from the DFT reference, provides a smooth and coherent optimization landscape; conversely, a more accurate hybrid pipeline that relies on an OOD-vulnerable surrogate may introduce calibration noise that destabilizes convergence. This observation shifts the evaluator design criterion from 'faithful mimicry of the reference' to 'preservation of monotonic gradients in the property-structure mapping'. When the evaluator's coverage is insufficient, even a capable reasoning engine may be misled by noisy feedback. This highlights an important direction for future AI4S frameworks: developing robust, adaptive evaluators that maintain accuracy across diverse chemical domains is essential for realizing the full potential of LLM-driven molecular design.

Further evidence that the model engages in reasoning lies in the fact that the RAG contained only a subset from the QM9 dataset, which comprises organic molecules with nine or fewer heavy atoms. Given that the HOMO-LUMO gap is influenced by the extent of conjugation, the QM9 dataset cannot be directly applied to tasks involving gaps of 3.0 eV or lower. The generated molecules (denoted as Figure 5) reveal that those with gaps of 3.0, 2.0 and 1.0 eV all contain more than nine heavy atoms. In contrast, for gaps of 4.0 and 5.0 eV, the generated molecules have fewer than nine heavy atoms, yet their structures remain distinct from those in the QM9 dataset. Notably, the molecule generated for the 5.0 eV gap task contains a sulfur atom —an element absent from the QM9 dataset—indicating that the model is capable of reasoning and can integrate the retrieved knowledge with its own world knowledge and reflection.

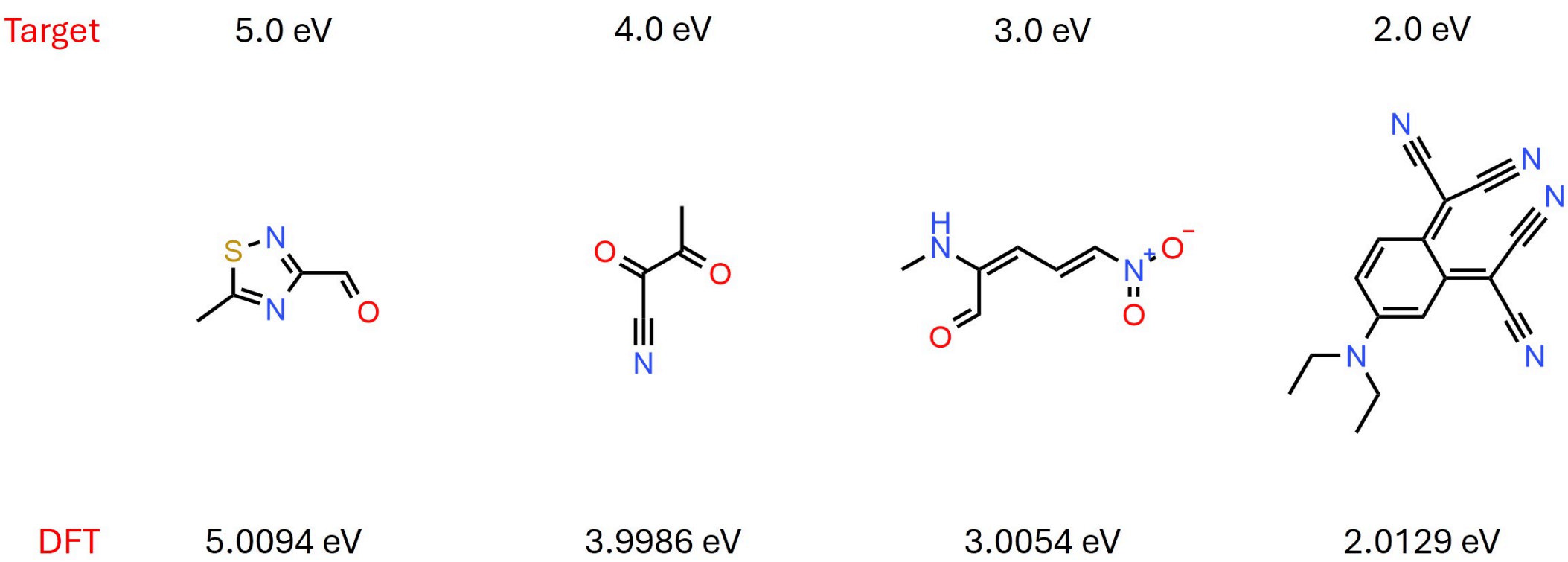

*Figure 5: Generated molecules of the SPR+RAG approach implemented by DeepSeek-V4Pro.*

## SPR vs scalar reflection

To determine the advantage of the SPR+RAG approach from the FORGE paradigm, we conducted a series of experiments on designing molecules with different DFT HOMO-LUMO gap targets with SPR+RAG, scalar+RAG (which only reflects the HOMO-LUMO gap itself), RAG-only (no reflection with RAG on), and baseline (no reflection and no RAG). The results were concluded in Table 2.

For simple tasks—such as targeting HOMO-LUMO gaps of 4.0 eV or 5.0 eV—the SPR+RAG approach outperforms other methods, achieving median and mean deviations of 0.0203 eV and 0.0174 eV, respectively, even though the baseline model alone already satisfies the tolerance threshold. For the moderate task with a 3.0 eV gap, the scalar+RAG approach shows comparable performance, yielding a smaller minimum deviation (0.0032 eV) but larger median and mean deviations (0.0108 eV versus 0.0061 eV, and 0.0117 eV versus 0.0068 eV, respectively). However, this performance comes at a higher computational cost than that of SPR+RAG.

Moreover, across all tasks, the scalar+RAG approach consistently consumes more tokens and time than SPR+RAG—a counterintuitive finding given that scalar reflections contain substantially less information than their SPR counterparts. This may be attributed to the fundamental difference between large language models and traditional deep learning models: less information may lead to a distinct reasoning process.

For the difficult task with a target gap of 2.0 eV, the SPR+RAG approach demonstrates both high performance and stability. The scalar+RAG approach again produces the lowest minimum deviation but higher median and mean deviations.

Overall, the RAG-only approach performed worse than the reflection-based methods, and its cost was comparable to theirs. Meanwhile, the baseline model (without RAG or reflection) also yielded poor performance but at drastically lower cost in terms of both time and token consumption. These results imply that a RAG system should always be paired with a reflection mechanism to realize its full potential. This finding has rarely been asserted in prior work, underscoring the significance of our observation.

## Batch reflection with pre-screening and per-molecule reflection

Theoretically, per-molecule reflection serves as the precise analog of a human researcher—albeit a single-threaded one. It mimics a depth-first search (DFS) in graph theory, where each molecule constitutes a node and each edge represents a possible transformation between two molecules. However, this approach can become trapped in a local minimum, with newly generated molecules remaining mere analogs of the initial compound.

In contrast, batch reflection resembles a breadth-first search (BFS). It not only extracts information from candidate molecules, but also explores the space of possible transformations horizontally. For example: “A contains a methyl group, and B contains a methoxy group. B has a higher HOMO, indicating that electron-donating ability can affect the HOMO energy.” By its very nature, batch reflection possesses the inherent capacity to extract structure-property relationships.

Our experimental results substantiate this claim. We performed a task designed to generate a molecule with a HOMO-LUMO gap of 3.0 eV, employing both the batch (each batch yields 20 candidates, and the top 5 are forwarded for DFT calculations) and per-molecule reflection mechanisms. The results are presented in Table 4. The batch strategy achieves a lower minimum deviation (0.0056 eV) than the per-molecule strategy (0.0157 eV), indicating that batch-level cross-candidate comparison can reach higher peak precision. Their median deviations are comparable (0.0137 vs. 0.0158 eV), suggesting that both strategies converge to similar typical performance. However, an important divergence emerges in their mean deviations: the per-molecule approach yields a lower mean (0.020 eV) than the batch strategy (0.0567 eV), and its standard error is an order of magnitude smaller (0.0043 vs. 0.0471 eV). This contrast reveals that while batch reflection can occasionally discover exceptionally precise solutions, its per-formance varies considerably across independent runs; the per-molecule strategy, by contrast, delivers more reproducible results. In other words, batch reflection trades consistency for peak performance—a characteristic consistent with BFS-like exploration, where broader sampling increases the chance of stumbling upon an optimal trajectory but also raises the risk of less favorable outcomes.

Regarding cost, the per-molecule approach requires less LLM time per iteration (56 s vs. 263 s), as expected given that it processes only one candidate per cycle. It also achieves a lower total LLM time (1128 s vs. 2631 s), although it consumes more tokens overall (117,714 vs. 54,550) because it runs for 20 iterations rather than 10.

Neither strategy is universally superior; rather, the choice between them entails context-dependent tradeoffs. When the design objective demands the highest achievable precision and the computational budget permits multiple independent runs to hedge against run-to-run variance, batch reflection with pre-screening is the preferred option. Conversely, when reproducibility and predictable convergence are paramount—as in high-throughput screening pipelines or resource-constrained deployments—the per-molecule strategy offers a more reliable alternative. In practice, the two approaches may also be combined: batch reflection for initial exploration of the chemical space, followed by per-molecule refinement to stabilize the final candidate.

| Mode | Success Rate | Min Dev (eV) | Median Dev (eV) | Mean±SE (eV) | LLM Time per Iteration (s) | LLM Time (s) | Tokens |
|---|---|---|---|---|---|---|---|
| Batch[a] | 3/3 | **0.0056** | **0.0137** | 0.0567±0.0471 | 263 | 2631 | **54,550** |
| Per-molecule[b] | 3/3 | 0.0157 | 0.0158 | **0.020±0.0043** | **56** | **1128** | 117,714 |

*Table 4: The performance of the batch reflection with pre-screening and per-molecule reflection in the design of molecules with a specific HOMO-LUMO gap of 3.0 eV (The experiments were performed in 16 threads parallel in a 2-way E5 2696 v4 workstation. [a] The batch strategy was executed in three independent runs, each comprising 10 iterations; [b] the per-molecule strategy was executed in three independent runs, each comprising 20 iterations).*

## Benchmark on different LLM models

Different LLMs exhibit varying capacities for generating valid molecules, a divergence attributable to their distinct training procedures, domain knowledge, and linguistic competencies. To evaluate the performance of different LLMs in the design of molecules with a specific HOMO-LUMO gap, we conducted a benchmark study on DeepSeek-V4Pro, DeepSeek-V4Flash, MiniMax-M3, QWen-3.7Max, and GLM5.1. The results are presented in Table 5.

| Rank | Model | Min Dev (eV) | Med Dev (eV) | Mean ± SE (eV) | LLM Time (s) | Tokens | Success Rate |
|---|---|---|---|---|---|---|---|
| 🥇 | DeepSeek v4-pro | 0.0054 | **0.0061** | 0.0238±0.0180 | 580 | 63,592 | 3/3 |
| 🥈 | Mimo v2.5-pro | 0.0053 | 0.0097 | 0.0250±0.0176 | 794 | 97,411 | 3/3 |
| 🥉 | Qwen 3.7-max | 0.0024 | 0.0112 | 0.0331±0.0265 | 628 | 90,489 | 3/3 |
| 4 | DeepSeek v4-flash | 0.0056 | 0.0137 | 0.0567±0.0471 | **200** | **54,550** | 3/3 |
| 5 | GLM 5.1 | 0.0030 | 0.0140 | 0.0315±0.0232 | 2999 | 159,742 | 3/3 |
| 6 | MiniMax M3 | 0.0103 | 0.0141 | **0.0141±0.0038** | 789 | 76,491 | 2/3 |
| 7 | Doubao seed-2-0-pro-260215 | **0.0004** | 0.0488 | 0.0509±0.0298 | 541 | 64,604 | 3/3 |

*Table 5: The performance of the FORGE paradigm in the design of molecules with a HOMO-LUMO gap of 3.0 eV. Models are ranked by median deviation across seeds.*

All seven models achieved deviations well below 0.1 eV, and all except MiniMax M3 (2/3) attained a 100% success rate, confirming that the SPR-guided framework generalizes effectively across diverse LLM backbones.

Ranking LLM performance on stochastic tasks requires careful metric selection. The minimum deviation captures the best-case outcome achievable through seed selection, but it is inherently optimistic and non-reproducible. The mean deviation summarizes expected performance but is sensitive to outlier seeds. The median deviation offers a robust middle ground: it reflects typical performance while resisting distortion from extreme values. We therefore rank models primarily by median deviation, but report all four statistics to enable task-specific interpretation.

By this criterion, DeepSeek v4-pro ranks first (median 0.0061 eV), achieving the lowest median deviation among all models together with a low standard error (0.0180 eV) and moderate LLM time (580 s)—a strong balance of precision, stability, and cost. Mimo v2.5-pro follows at second (median 0.0097 eV) with the smallest max deviation (0.0601 eV), indicating excellent reproducibility. Qwen 3.7-max holds third (median 0.0112 eV) with the best min deviation (0.0024 eV), though its higher max deviation indicates greater inter-seed variability.

Doubao seed-2-0-pro-260215 presents the most striking divergence between metrics: it achieves the lowest minimum deviation of any model (0.0004 eV)—an exceptionally precise single result—yet ranks last by median (0.0488 eV) and mean (0.0509 eV). Its per-seed breakdown (0.0488, 0.0004, 0.1034) reveals that the best result is a rare outlier flanked by a mediocre seed and a severely divergent one (max 0.1034 eV). This pattern suggests that Doubao is capable of extraordinary precision but lacks the consistency to deliver it reliably.

At the opposite extreme, MiniMax M3—despite the worst minimum deviation (0.0103 eV) and incomplete reproducibility (2/3 seeds)—exhibits the lowest standard error (0.0038 eV) and the smallest max deviation (0.0179 eV) among all models. Its narrow deviation range (0.0103-0.0179 eV) indicates that when it succeeds, its performance is highly predictable, albeit at a moderate precision level.

DeepSeek v4-flash (median 0.0137 eV) and GLM 5.1 (median 0.0140 eV) occupy the middle tier with comparable typical performance, although GLM exhibits lower variance (SE 0.0232 vs. 0.0471 eV). DeepSeek v4-flash demonstrates high cost-efficiency, consuming the least LLM time (200 s) and fewest tokens (54,550)—an order of magnitude faster than most competitors —making it the most economical choice when sub-0.02 eV precision suffices.

These results underscore that minimum deviation alone is an insufficient basis for model selection in molecular design. A model that achieves near-perfect results on one seed but large errors on others (e.g., Doubao) may be less practically useful than one with slightly higher but more consistent deviations (e.g., Mimo). We therefore recommend reporting median deviation as the primary ranking metric, supplemented by min, mean ± SE, and median deviations for a complete characterization of model behavior.

## Performance on different target properties

As the strategy shows promising performance in designing molecules with specific HOMO-LUMO gaps, we are going to evaluate its capability in other properties to confirm its versatility in this subsection.

### Dipole moment

The dipole moment is one of the important properties in the design of molecules, as it can affect the electronic properties of the molecule and the interaction with the environment. Designing molecules with a specific dipole moment is a common task in the field of materials science. We tested our SPR relection cycle on designing with a specific dipole moment of 2.5 D. The results were shown as Table 6.

| Target Dipole Moment (D) | Min Dev (D) | Median Dev (D) | Mean ± SE (D) | LLM Time (s) |
|---|---|---|---|---|
| 2.5 | 0.016 | 0.038 | 0.078±0.073 | 1139 |

*Table 6: The performance and costs of different LLM models in the design of molecules with a specific dipole moment of 2.5 D (LLM: DeepSeek-V4Flash, no thinking).*

Figure 6 shows the best candidate from each iteration across 3 independent runs. As can be seen, the final candidates were not always generated in the final iteration. Once the system had reached a small deviation that approached the boundaries of the model, further reflection conversely led to both worse results and oscillating deviations. This trend is consistent with our previous results for the HOMO-LUMO gap.

This first run serves as a useful example of how the model generates molecules with a specific dipole moment. Initially, the best candidate was bromobenzene, which has a dipole moment of 1.6304 D—a deviation of 0.8696 D from the target value of 2.5 D. The model then attempted to replace the halogen atom (bromine, a negative center) with a more positive substituent, namely an acetyl group. However, the dipole moment of acetylbenzene was 3.0211

D, exceeding the target. Next, the model introduced a para-methoxy group to form a typical donor-acceptor structure, but this did not reduce the deviation. Following this iteration, a local oscillation emerged between 4-methoxy-acetylbenzene and 4-methyl-benzaldehyde. In the sixth iteration, the model reached its best candidate, which is essentially a combination of the two preceding structures: 4-methyl-3-methoxy-benzaldehyde. This compound has a dipole moment of 2.5164 D, the closest among all candidates. Subsequent efforts to further reduce the deviation all failed. Compared to a human researcher, the model exhibits even greater and more precise chemical intuition in predicting how the dipole moment will change upon the introduction of a new functional group.

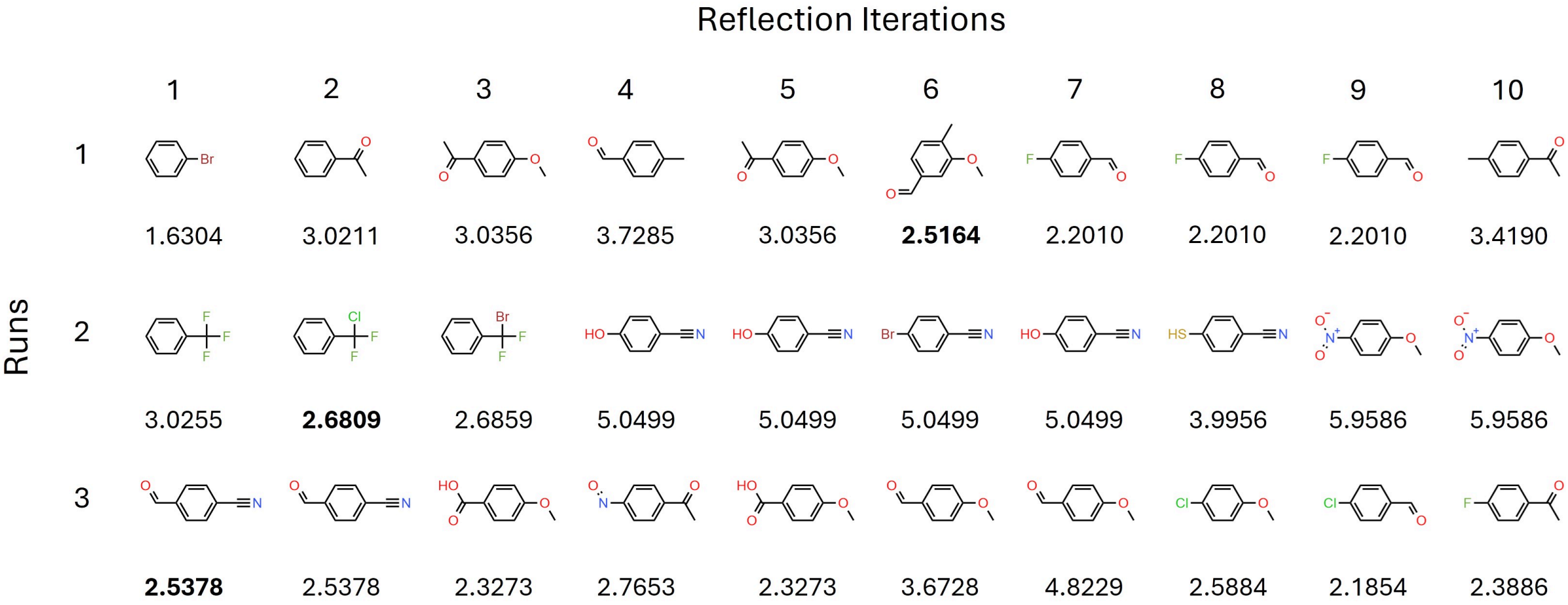

*Figure 6: The candidate that best matched the target dipole moment of 2.5 D from each iteration across three independent runs. The final set of candidates comprised those exhibiting the smallest deviation from the target; these are highlighted in boldface (LLM: DeepSeek-V4Flash, no thinking).*

## Conditional optimizing LogP of aggregation-induced emission molecules

Aggregation-induced emission (AIE) systems represent an emerging class of light-emitting materials that exhibit strong emission in aggregated states but weak or negligible emission in dilute solution—a behavior that stands in stark contrast to conventional fluorescent materials, which show intense emission in dilute solution but markedly diminished or no emission upon aggregation.[27,28] In recent years, AIE systems have found prominent applications in biomedicine, serving as imaging agents, components for photodynamic therapy, and beyond. However, a defining structural feature of AIE systems is their flexible aromatic skeleton, which is inherently hydrophobic, thus limiting their direct applicability in pharmaceutical contexts. The conventional approach involves using amphiphilic molecules to encapsulate AIE dyes into nanovesicles—a strategy that can complicate the formulation process and introduce uncertainty in practical applications.

LogP is one of the key properties in the design of drug molecules. According to Lipinski's rule of five, an ideal drug candidate should have a LogP value less than 5, typically ranging from 2-4, or even 1-3.[29] The most archetypal AIE molecule, tetraphenylethylene (TPE), comprises four benzene rings and possesses a LogP of 6.69, indicating that TPE is highly unsuitable for biological applications.

To address this limitation, we introduce FORGE as a proof-of-concept method for reducing the LogP of AIE molecules (exemplified by tetraphenylethylene, TPE) while preserving their core structural motifs and electronic properties, such as the HOMO-LUMO gap. The modification to the previously employed architecture is straightforward: we simply set TPE as the initial structure and introduce three optimization objectives — LogP (target), deviation of the HOMO-LUMO gap (condition), and Tanimoto similarity (structural constraint) — while keeping other available reflection information.

We performed 20 iterations of SPR+RAG reflection on TPE to optimize its LogP while preserving its HOMO-LUMO gap and structural skeleton. Although this experiment employed the XTB-computed gap as a reference to improve efficiency, the DFT-computed gap can also be readily integrated into the same framework.

The LogP and the GFN2-XTB-level HOMO-LUMO gap of TPE are 6.69 and 3.39 eV, respectively. For each iteration, the model was instructed to generate 20 candidates, which were then optimized with XTB. Their LogP values were computed using RDKit, and additional reflection information was extracted, including the HOMO-LUMO gap, the HOMO and LUMO energies, and the Tanimoto similarity of the Morgan fingerprint relative to TPE. Every five iterations constituted one phase. In the subsequent phase, the molecule with the lowest combined value of LogP and HOMO-LUMO gap deviation from the previous phase was adopted as the starting structure. A total of four phases were performed.

The results were shown in Table 7. A candidate is deemed compliant if it satisfies both of the following constraints relative to the phase baseline:

- The LogP change exhibits the correct sign — if the objective is to reduce LogP, any negative ΔLogP (even −0.01) is accepted. No threshold on the magnitude is imposed.
- The gap drift does not exceed the tolerance (0.5 eV)

This criterion is deliberately lenient: it merely verifies that the modification did not worsen the properties. In the first phase, the LogP decreased by 0.75, with a gap drift of 0.2450 eV. Over subsequent iterations, the LogP continued to decline, while the gap drift remained within the tolerance. The optimal candidate emerged during the third iteration of the fourth phase. Its LogP was 3.26 — substantially below the Lipinski threshold and approaching the ideal value of 3.0 — and the gap drift was 0.4947 eV.

| Phase | Iteration | LogP | Δ LogP | Gap Drift (eV) | Compliants | tokens | LLM Time (s) |
|---|---|---|---|---|---|---|---|
| 1 | 5 | 5.94 | 0.75 | 0.2450 | 8/20 | 4093 | 45 |
| 2 | 3 | 5.65 | 1.04 | 0.3434 | 2/20 | 4296 | 51 |
| | 4 | 4.76 | 1.93 | 0.1604 | 13/20 | 4357 | 46 |
| | 5 | 5.06 | 1.63 | 0.4204 | 11/20 | 2412 | 20 |
| 4 | 2 | 4.01 | 2.68 | 0.3481 | 4/20 | 3218 | 32 |
| | 3 | 3.26 | 3.43 | 0.4947 | 8/20 | 2324 | 21 |

*Table 7: The performance and costs of the structural optimization on TPE for lower LogP values (LLM: DeepSeek-V4Flash).*

**SA score optimization when keeping specific properties**

The synthetic accessibility (SA) score is a metric that quantifies the ease of synthesizing a molecule.[30] It was originally employed to pre-screen chemicals from large ensembles. In practice, however, this approach requires a sufficiently large ensemble for effective pre-screening—a prerequisite that is not always met. For instance, if one has already designed a molecule with desirable properties (e.g., a favorable $S_0$-$S_1$ excitation energy) yet encounters synthetic difficulties, relying on the SA score to guide structural modifications toward simpler synthesis becomes challenging, because there is no ensemble containing molecules analogous to the initial design.

However, FORGE offers a valuable alternative. Using the initial design as a starting point, one can perform SPR+RAG reflection to optimize the SA score without the need for a large ensemble of analogues. The LLM generates a series of analogues derived from the initial structure, which are then ranked based on both their SA scores and the deviation from the target property.

This is also a proof-of-concept experiment. Our initial structure is a N,N′-diethyl-1,2,3,4-tetrahydro-quinoxaline-based coumarin, which is a typical fluorscence dyes.[31] However, the N,N′-diethyl-1,2,3,4-tetrahydro-quinoxaline structure is of a little bit synthesis difficulty (2.596 of the SA score). So the experiment could demonstrate the potential of the FORGE system to optimize the structuer by keeping the S0-S1 excitation energy (3.37 eV of the $S_0$-$S_1$ excitation energy). The evaluator presented herein was developed using RDKit, XTB, and ORCA rather than pySCF, as ORCA exhibits superior efficiency in TD-DFT calculations (PBE0/def2-SVP with nstates = 10) when the RIJCOSX approximation is enabled.[32–34]

The results are presented in Figure 7. To illustrate the reasoning process, we highlight the best candidates. The most synthetically challenging moiety in the initial structure is the N,N′-diethyl-1,2,3,4-tetrahydro-quinoxaline ring. After several iterations, the LLM generated a structure at iteration 6 in which this ring was replaced by a simpler N,N-diethyl amino group. However, the electron-donating ability of the N,N-diethyl group is substantially weaker than that of the N,N′-diethyl-1,2,3,4-tetrahydro-quinoxaline ring. Consequently, although the SA score decreased to 2.135, the $S_0$-$S_1$ excitation energy dropped to 3.871 eV, corresponding to a deviation of 0.51 eV from the target value.

To compensate for the reduced electronic effect, the model introduced a cyano group as an electron-withdrawing substituent. This modification increased the SA score, but the deviation in the $S_0$-$S_1$ excitation energy was reduced to 0.21 eV. In iteration 9, the structure from iteration 7 was refined further, lowering the SA score to 2.280 while maintaining the $S_0$-$S_1$ excitation energy at 3.582 eV—a deviation of only 0.22 eV from the initial structure. Remarkably, this process closely mirrors the iterative reasoning of a human researcher, yet the total runtime amounts to only around 600 seconds — of which 248 seconds are attributable to LLM inference — with a token consumption of 17,017.

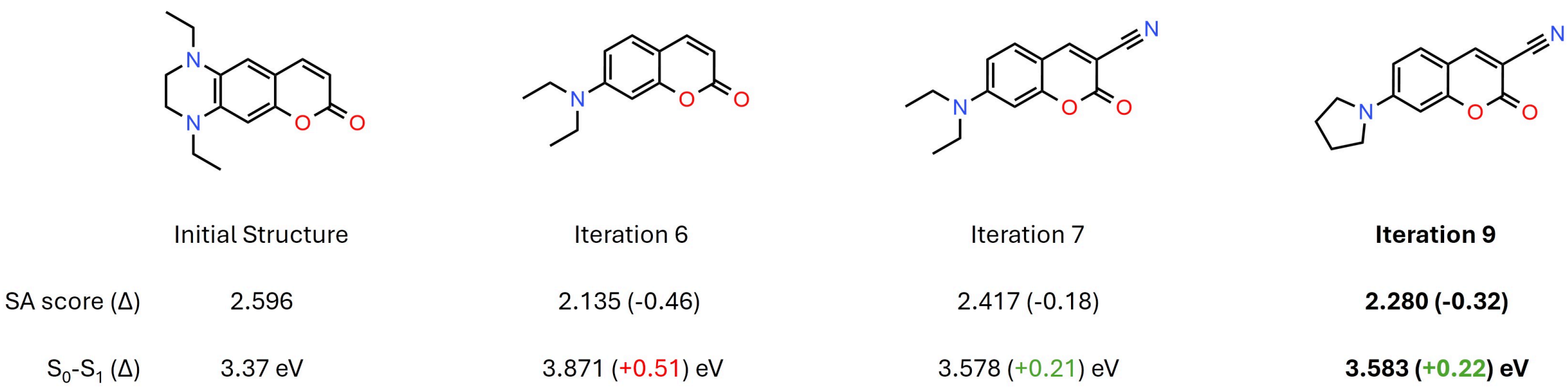

*Figure 7: The best candidates of each iteration with lowest SA score and the HOMO-LUMO gap shifts.*

### Docking optimization on a target protein

Docking optimization is a routine task in drug discovery, aiming to identify the most stable binding pose of a molecule with its target protein. Most LLM-based molecular generators are designed for this purpose. To verify that FORGE constitutes a universal paradigm rather than a domain-specific system, we performed an experiment demonstrating how FORGE can optimize the docking ability of methotrexate to dihydrofolate reductase (DHFR).

The procedure comprises three independent runs, each consisting of ten rounds of reflection. In each run, the LLM generates a series of analogues derived from the initial structure, and the evaluator ranks them according to both their docking scores and the deviation from the target property. The reflection information includes the molecular structure, docking score (absolute value and difference from methotrexate), and contact information (observed via ProLIF, e.g., anionic contacts (ARG38), van der Waals contacts (SER59), and counts of different contact types). In the absence of explicit electronic-structure descriptors within the scoring function, these interaction fingerprints serve as the mechanism-level explanation required by SPR: they inform the LLM precisely which intermolecular contacts—such as a missing salt bridge or an under-exploited hydrophobic pocket—can be modulated through structural modification, thereby preserving the analytical logic of the framework even in a classical docking context.

In detail, all molecular docking was performed with Smina[35] using the Vinardo scoring function[36]. Vinardo was selected over the default Vina scoring function due to its superior ranking performance in retrospective virtual screening benchmarks, with a median enrichment factor at 1% (EF1%) of 44.6 versus 29.0 for Vina across 102 targets in the DUD-E dataset. It should be noted that Vinardo systematically assigns less negative scores than Vina. Therefore, the absolute docking scores reported here should not be directly compared with absolute values from studies using the Vina scoring function; however, the intra-target ranking of compounds in the present work is reliable.

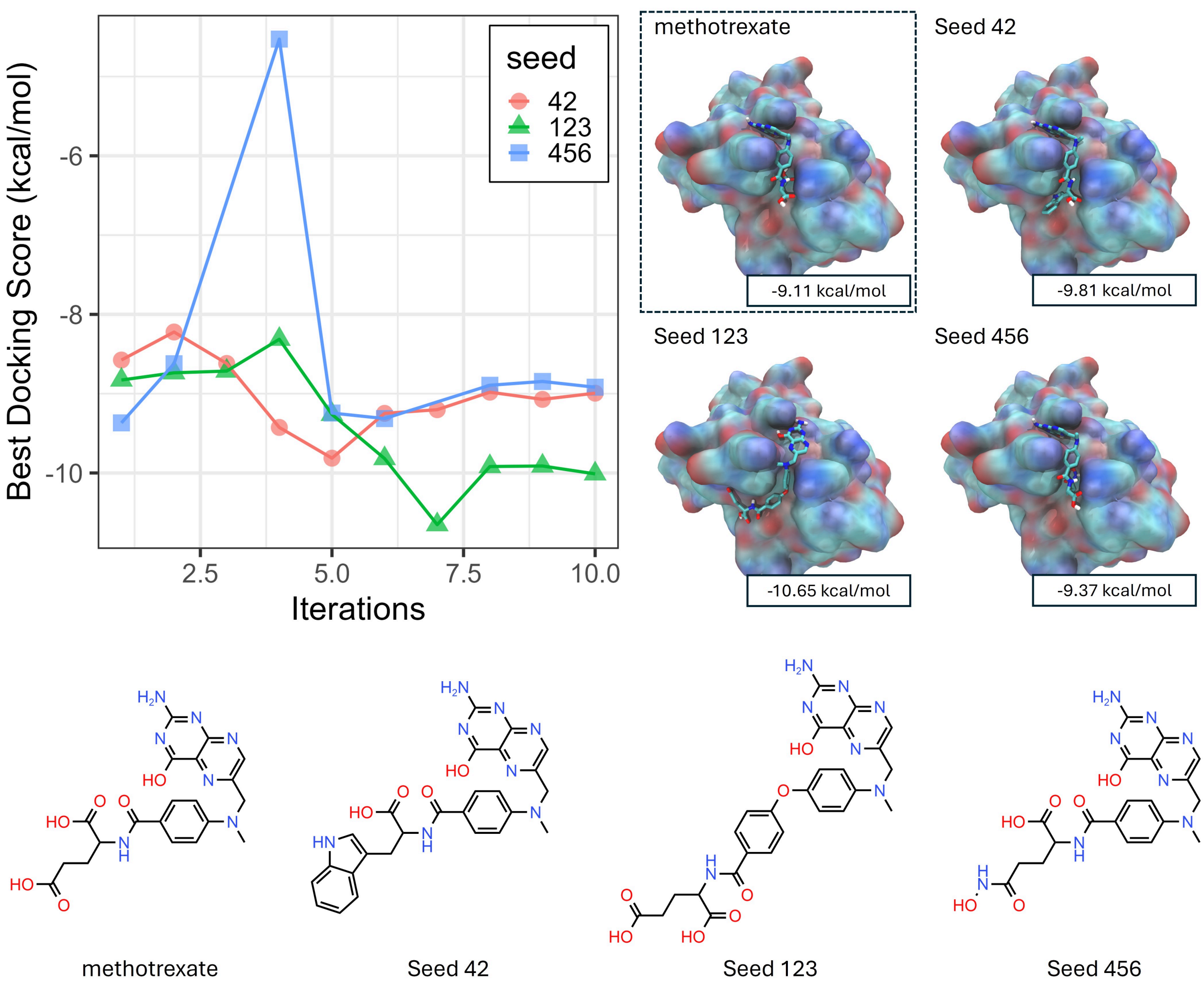

*Figure 8: The plots of the variation of the best docking score over three independent runs and the best pose and chemical structure (lowest-scoring candidate) from each run.*

The results of the docking optimization are presented in Figure 8. All experiments were completed in 33.7 min, requiring 816 s of LLM computation time and consuming 139,710 tokens in total. Compared to the best pose of methotrexate, which had a docking score of −9.11 kcal/mol, each run yielded improved results within 10 iterations. The best docking score was approximately −10.65 kcal/mol across all three independent runs, and the best pose from each run is shown in the figure. The skeleton of methotrexate is pteridine—Ph—C(=O)NH—Glu(COOH). To improve its docking score, seed 42 attempted to replace the —C(=O)NH—Glu(COOH) moiety with —C(=O)NH—indole, which enhanced the van der Waals contacts and introduced a new interaction (VdW·ASN64). The best candidate from seed 123 replaced the —C(=O)NH—Glu(COOH) moiety with —O—Ph—C(=O)NH—Glu(COOH), which drastically enhanced the van der Waals contacts and introduced a strong interaction (VdW·TYR22).

## Limitations and outlook

For rapid in situ evaluation, our implementation constructs the evaluator based on in silico computed properties. However, this paradigm is not limited to such an approach. In principle, the system can be adapted to evaluate molecules using other criteria—including experimental data, pre-trained machine-learning predictors or regressors, or even human feedback. Nevertheless, with efficiency as a primary consideration, the in silico evaluator is best positioned to realize the full potential of the system.

# Conclusion

This work began with a simple question: what happens when we stop telling a language model whether a molecule is good or bad, and instead present it with the full quantum-mechanical evidence for why? The answer, demonstrated across 4 HOMO-LUMO gap targets, dipole-moment design, LogP and synthetic-accessibility optimization, molecular docking, and 7 LLM backbones, is that the model ceases to guess and begins to reason. The SPR-guided reflection loop consistently achieves sub-0.02 eV median deviations across all tested targets, while scalar feedback becomes markedly less competitive on the most challenging tasks. Beyond molecular design, our results point to a broader principle for scientific AI: iterative systems that close the loop with mechanism-level explanation—rather than outcome-level scores—can bridge the gap between prior knowledge and posterior insight, transforming large language models into genuine partners in discovery.

# Code availability

The code for our work is available at https://github.com/JGong-CatenaryGong/forge.